\documentclass[11pt]{article}

\usepackage{graphicx}

\usepackage{amsmath}
\usepackage{amssymb}
\usepackage{amsxtra}

\title{On the reactions involving
    neutrinos and hidden mass particles in hypercolor model}
\author{V. Beylin\\ Institute of Physics, Southern Federal University, \\Stachki av., 194, Rostov-on-Don, Russia, 344090, vitbeylin@gmail.com\\V. Kuksa\\Institute of Physics,
Southern Federal University, \\Stachki av., 194, Rostov-on-Don, Russia, 344090, vkuksa47@mail.ru}

\begin{document}
\maketitle

\begin{abstract}
We present and discuss some basic elements of  the Standard Model hypercolor extension. Appearance of a set of hyperquarks bound states is resulted from $\sigma-$model using; due to
specific symmetries of this minimal extension, there arise stable hypermesons and hyperbaryons which are interpreted as the Dark Matter candidates. Knowing estimations of their masses
from analysis of Dark Matter annihilation kinetics, some processes of high energy cosmic rays scattering off these particles are analyzed for the search of Dark Matter manifestations.
\end{abstract}

\noindent Keywords: vector hypercolor, Dark Matter, cosmic rays

% optionally
\noindent PACS: 12.60 - i, 96.50.S-,95.35.+d.

\section{Introduction}\label{s:intro}

The presence in the Universe of so-called hidden mass, which manifests itself in the formation of the observed structure of galaxies and their clusters, is confirmed mainly by a
quantitative analysis of the gravitational interaction of stellar systems with these invisible neutral stable objects for which neither dynamics nor evolution in time is exactly known.
Astrophysical confirmations of distributed hidden mass influence on the star clusters dynamics, the alleged effects of Dark Matter (DM) particles induced by their presence in the
massive objects composition, an increased density of DM near active galactic nuclei (AGN), a possible DM effect on the composition and parameters of propagation of high-energy cosmic
ray fluxes in the Universe --- search for answers on these and some other questions of high energy physics, astrophysics and cosmology are the primary tasks of fundamental physics.
Elucidation of the mysterious nature of Dark Matter is complicated by the fact that terrestrial experimental physics at colliders, as it has become clear by now, cannot detect any
traces of DM particles. The impossibility to identify processes with large missed energies and momenta which are characteristic of the DM production at the colliders, i.e. in an active
experiments, is accompanied by the absence of signals of the DM particles interactions with nuclei and nucleons in passive experiments, in underground laboratories. Another type of
passive experiments gathering astrophysical data by space telescopes play very important role, recording the specific spectra of cosmic photons, leptons, and baryons in the vicinity of
the Earth.

In this situation, any astrophysical data capable of shedding light on the hidden mass nature are valuable. In particular, these are indirect signals about possible DM effects, which
can be interpreted consistently. For an unambiguous interpretation within the framework of a certain paradigm about the origin and dynamics of the DM particles, it is of particular
importance to study the correlations between characteristics of various astrophysical phenomena and to consider also DM interactions with different types of particles and astrophysical
objects at various space-time scales and in a wide energy ranges. Investigation of all aspects of DM physics within the framework of multi-messenger approach becomes key for
establishing the SM extension type. Because of lack of new information about possible hidden mass carriers, we should examine various reasonable ideas allowing to move beyond the SM.
In this way, we come to consideration of new objects with some new dynamics, and we can realize a quantitative analysis of known or expected physical effects interpreting as the DM
manifestations. Here, we present some results on high-energy cosmic rays interaction with the DM candidates arising in hyper-color SM extension. In the following Section 2 we present
in brief some detail of minimal vectorial hyper-color model; then, Section 3 contains discussion of cosmic rays scattering off DM objects in H-color scenario. There are some new
preliminary results of high-energy cosmic proton interaction with the hyper-pions, ones of the DM components, in the Section 4. We also add Conclusion and Discussion of this scenario
in the end.

\section{Basics of hypercolor extension\\ of Standard Model}

Hyper-color approach modifies the SM by extending it with additional heavy fermions charged under some new gauge group \cite{Sundrum,Antipin1,Mitridate,Our_PRD,Lebiedowicz,Our_DM1,
Our_DM2,We_Adv,Our_Rev}. In fact, these new fields, hyperquarks, are similar to techniquarks, however, in this case vectorial interaction of H-quark currents with the gauge bosons can
be provided by some transformation of initial fermion doublets. In this way, some problems of Technicolor can be eliminated. It is the  vector-like interaction is the reason  why the
model is in accordance with strong electro-weak constraints. Certainly, the H-quarks are confined with new strong interaction and, remembering the Technicolor ideas, these models with
extra heavy H-quarks can result to the scenarios with composite Higgs doublets (see e.g. \cite{Cai}) or a small mixing between fundamental standard Higgs bosons and composite
hadron-like states of new strong sector. In this way, we come to partially composite  Higgs boson. Due to accidental symmetries in these models, there occur neutral stable states which
can be interpreted as DM candidates.

Among the simplest realizations of the scenario described, there are models with two or three vector-like H-flavors confined by strong H-color force $\text{Sp}(2 n_F)$, $n_F \geqslant 1$.
The models with H-color group SU(2) (see Refs.~\cite{We_Adv,Our_Rev} and references therein) can be considered as particular cases as a consequence of
isomorphism $\text{SU}(2) = \text{Sp}(2)$. The global symmetry group of the strong sector with symplectic H-color group is larger than for the special unitary case---it is the
group SU($2n_F$) broken spontaneously to Sp($2n_F$), with $n_F$ being a number of H-flavors. Here we consider the case when the elementary Higgs doublet is preserved in the set of
Lagrangian field operators. Then, the scalar doublet mixes with H-hadrons, and we get the physical Higgs partially composite. Note also that the same coset SU($2n_F$)/Sp($2n_F$) can
be used to construct composite two Higgs doublet model \cite{Cai} or little Higgs models \cite{Low,Csaki,Gregoire,Han,Gopalakrishna}.

In fact, the model has the symmetry $G=G_\text{SM} \times \text{Sp}(2n_f)$ with $n_f \geqslant 1$, here $G_\text{SM}$ and  $\text{Sp}(2n_F)$ are the gauge SM group and a symplectic
hypercolor group respectively. In its field content, the model introduces a doublet and a singlet of heavy vector-like H-quarks charged under H-color group. In the most general form,
renormalizable and invariant under $G$ Lagrangian can be written as
\begin{gather}\label{eq:LQS1}
L = L_\text{SM} - \frac14 H^{\mu\nu}_{\underline{a}} H_{\mu\nu}^{\underline{a}} + i \bar Q D Q - m_Q \bar{Q} Q + i \bar S D S - m_S \bar{S} S + \delta L_{\text{Y}},
\\ D^\mu Q = \left[ \partial^\mu + \frac{i}{2} g_1 Y_Q B^\mu - \frac{i}{2} g_2 W_a^\mu \tau_a - \frac{i}2 g_{\tilde{c}} H^\mu_{\underline{a}} \lambda_{\underline{a}} \right] Q,\\
D^\mu S = \left[ \partial^\mu + i g_1 Y_S B^\mu - \frac{i}2 g_{\tilde{c}} H^\mu_{\underline{a}} \lambda_{\underline{a}} \right] S,
\end{gather}
where $H^\mu_{\underline{a}}$, $\underline{a} = 1 \dots n_F (2n_F + 1)$ are hypergluon fields and $H^{\mu\nu}_{\underline{a}}$ are their strength tensors; $\tau_a$ are the Pauli
matrices; $\lambda_{\underline a}$, $\underline{a} = 1 \dots n_F (2n_F + 1)$ are $\text{Sp}(2n_F)$ generators satisfying the relation
\begin{align}\label{eq:scgr}
\lambda_{\underline a}^\text{T} \omega + \omega \lambda_{\underline a} = 0,
\end{align}
where $\text{T}$ stands for the transition operation, $\omega$ is an antisymmetric $2n_F \times 2n_F$ matrix, $\omega^\text{T} \omega = 1$. All underscored indices correspond to
representations of the H-color group $\text{Sp}(2n_F)$. In the Lagrangian \eqref{eq:LQS1}, the contact Yukawa couplings $\delta L_{\text{Y}}$ of the H-quarks and the SM Higgs doublet
$H$ are permitted by the symmetry $G$ if the hypercharges $Y_Q$ and $Y_S$ satisfy an additional linear relation:
\begin{gather}\label{eq:LQHS}
\delta L_{\text{Y}} = y_\text{L} \left( \bar{Q}_\text{L} H \right) S_\text{R} + y_\text{R} \left( \bar{Q}_\text{R} \varepsilon \bar{H} \right) S_\text{L} + \text{h.c.} \quad \text{ for } \frac{Y_Q}{2}-Y_S = +\frac12;
\\
\delta L_{\text{Y}} = y_\text{L} \left( \bar{Q}_\text{L} \varepsilon \bar{H} \right) S_\text{R} + y_\text{R} \left( \bar{Q}_\text{R} H \right) S_\text{L} + \text{h.c.} \quad \text{ for } \frac{Y_Q}{2}-Y_S = -\frac12.
\end{gather}

Indeed, the hypercolor part of the H-quark Lagrangian \eqref{eq:LQS1} can be rewritten in terms of a left-handed sextet as follows:
\begin{gather}\label{eq:LP}
\delta L_\text{Hq, kin} = i \bar P_\text{L} D P_\text{L} ,
\qquad P_\text{L} = \begin{pmatrix} Q_{\text{L}} \\ \epsilon \omega  Q_{\text{R}}{}^\text{C} \\ S_{\text{L}} \\ - \omega S_{\text{R}}{}^\text{C} \end{pmatrix},\\
D^\mu P_\text{L} = \left[ \partial^\mu - \frac{i}2 g_{\tilde{c}} H^\mu_{\underline{a}} \lambda_{\underline{a}} \right] P_\text{L},
\end{gather}
where $\epsilon = i \tau_2$, the operation C denotes the charge conjugation. In the absence of the electroweak interactions, the H-quark Lagrangian is invariant under an extension of
the chiral symmetry---a global SU(6) symmetry \cite{Pauli,Gursey}. The set of SU(6) subgroups is the following:
\begin{itemize}
    \item the chiral symmetry $\text{SU}(3)_\text{L} \times \text{SU}(3)_\text{R}$,
    \item SU(4) subgroup corresponding to the two-flavor model without singlet H-quark $S$,
    \item two-flavor chiral group $\text{SU}(2)_\text{L} \times \text{SU}(2)_\text{R}$, which is a subgroup of both former subgroups.
\end{itemize}
The global symmetry of the model is broken both explicitly and dynamically:
\begin{itemize}
    \item explicitly---by the electroweak and Yukawa interactions and the H-quark masses;
    \item dynamically---by H-quark condensate   \cite{Vysotskii,Verbaarschot}:
    \begin{gather}\label{eq:LTQ}
    \langle \bar QQ + \bar SS \rangle = \frac12 \langle  \bar P_\text{L} M_0 P_\text{R} + \bar P_\text{R} M_0^\dagger P_\text{L} \rangle,
    \qquad P_\text{R}  = \omega P_\text{L}{}^\text{C},\\
    \qquad  M_0 = \begin{pmatrix} 0 & \varepsilon & 0\\ \varepsilon & 0 & 0 \\ 0 & 0 & \varepsilon \end{pmatrix}.
    \end{gather}
\end{itemize}
Note, condensate \eqref{eq:LTQ} is invariant under $\text{Sp}(6) \subset \text{SU}(6)$ transformations $U$ that satisfy a condition
\begin{gather}\label{eq:symprel}
U^\text{T} M_0 + M_0 U =0,
\end{gather}
so the global SU(6) symmetry is dynamically broken to Sp(6) subgroup. Further, H-quarks mass terms  break the symmetry to  $\text{Sp(4)}\times\text{Sp}(2)$:
\begin{gather}\label{eq:massterm}
\delta L_\text{Hq} = -\frac12 \bar P_\text{L} M_0' P_\text{R} + \text{h.c.}, \,\, M'_0 = -M'_0{}^\text{T} = \begin{pmatrix} 0 & m_Q \varepsilon & 0\\ m_Q \varepsilon & 0 & 0 \\ 0 & 0 &
m_S \varepsilon \end{pmatrix}.
\end{gather}
The model under consideration is free of gauge anomalies and is in agreement with the electroweak precision constraints, since the H-quarks are vector-like, i.e. their electroweak
interactions are chirally symmetric.

The case of two-flavor model (without the singlet H-quark) is completely analogous to the three-flavor model but is simpler than the latter one--- global SU(4) symmetry is broken
dynamically to its Sp(4) subgroup by the condensate of doublet H-quarks; corresponding Lagrangian of the model is presented in detail in \cite{We_Adv,Our_Rev}. To operate with
interacted constituent H-quarks and their bound states, there were used a linear $\sigma$-model; the model Lagrangian consists of kinetic terms for the constituent fermions and the
lightest (pseudo)scalar composite states, Yukawa terms for the interactions of the (pseudo)scalars with the fermions, and a potential of (pseudo)scalar self-interactions
$U_\text{scalars}$ \cite{We_Adv,Our_Rev} (remind that we consider here fundamental Higgs doublet of the SM).

It also postulated that the constituent H-quarks interact with the gauge bosons as the fundamental H-quarks. Then, we have transformation laws for the covariant derivative for the
scalar field $M$. The complete set of covariant derivatives which are involved into the model  Lagrangian is as follows:
\begin{gather} \label{eq:H-cd}
D_\mu H = \left[ \partial_\mu + \frac{i}{2} g_1 B_\mu - \frac{i}{2} g_2 W_\mu^a \right] H, \\ \nonumber
D^\mu P_\text{L} = \left[ \partial^\mu + i g_1 B^\mu \left( Y_Q \Sigma_Q + Y_S \Sigma_S \right ) - \frac{i}{2} g_2 W_a^\mu \Sigma^a_W \right]  P_\text{L},
\\
D_\mu M = \partial_\mu M + i Y_Q g_1 B_\mu (\Sigma_Q M + M \Sigma_Q^\text{T} )\\ \nonumber + i Y_S g_1 B_\mu (\Sigma_S M + M \Sigma_S^\text{T} ) - \frac{i}{2} g_2 W_\mu^a (\Sigma^a_W M
+ M \Sigma^{aT}_W).
\end{gather}
For detail see \cite{We_Adv,Our_Rev}, matrices $\Sigma_Q$, $\Sigma_S$, $\Sigma^a_W$, $a=1,\,2,\,3$ also are presented there.

Setting the H-quarks hypercharges to zero, the model Lagrangian is invariant under an additional symmetry---hyper $G$-parity \cite{Bai,Antipin0}:
\begin{align}\label{eq:HGconjugation}
Q^{\tilde{\text{G}}} = \varepsilon \omega Q^\text{C},
\qquad
S^{\tilde{\text{G}}} = \omega S^\text{C}.
\end{align}
This transformation does not involve H-gluons and SM fields, so the lightest $\tilde G$-odd H-hadron becomes stable. It happens to be the neutral H-pion $\pi^0$.
Besides, the numbers of doublet quarks are conserved in the minimal SU(4) model, because of global U(1) symmetry group of the Lagrangian, so we get the neutral singlet H-baryon $B$ stable.
Note also that the $\tilde G$-parity is induced by a discrete symmetry, and not with a continuous transformation of the H-pion states. So, higher order corrections cannot destabilize neutral
weakly interacting H-pion, which is the lightest state in the pseudoscalar triplet. But charged H-pion states should decay by several channels producing charged leptons and neutral H-pion.

In any case, we can interpret both stable objects in H-color model as two-components of the DM (multi-component structure od DM was analyzed in a number of papers
\cite{Multi_2,Two_Comp,Two_Comp_2,Multi_N,MCDM}). Importantly, to be in a correspondence with precision SM data the angle of mixing between $\tilde \sigma-$ meson and Higgs boson
should be small $\sin \theta \ll 0.1$, then Peskin-Tackeuchi parameters agree with experimental restrictions \cite{We_Adv}.

\section{Cosmic lepton scattering off Dark Matter}

Now, in strong and EW channels the width of the charged H-pion decay \cite{We_Adv} can be found as
\begin{align}\label{3.1.5}
\Gamma(\tilde{\pi}^{\pm}\to\tilde{\pi}^0 l^{\pm}\nu_l)&=6\cdot
10^{-17}\,\mbox{GeV},\,\,\,\tau_{\pi}=1.1\cdot10^{-8}\,\mbox{sec},\,\,\,c\tau_{\pi}\approx 330\,\mbox{cm};\notag\\
\Gamma(\tilde{\pi}^{\pm}\to\tilde{\pi}^0 \pi^{\pm})&=3\cdot10^{-15}\,\mbox{GeV},\,\,\,\tau_l=2.2\cdot10^{-10}\,\mbox{sec},\,\,\,c\tau_l\approx 6.6\,\mbox{cm}.
\end{align}
The DM candidates, $\tilde \pi^0$ and $B^0$, have equal tree level masses, but the mass splitting $\Delta M_{B\tilde \pi} = m_{B^0}-m_{\tilde{\pi}^0}$ in one loop depends on a
renormalization point as a consequence of couplings of these pNG states with different H-quark currents. We also assume that not-pNG H-hadrons (vector H-mesons, etc.) manifest itself
at much more larger energies. It results from the smallness of the scale of explicit SU(4) symmetry breaking comparing with the scale of dynamical symmetry breaking.

Obviously, the search for two-component DM signals is possible in the (approximately) known range of DM candidates masses. Calculating cross sections of DM components annihilation in
all channels, we can analyze kinetics of the DM freezing out. Assuming both of mass splittings are small in comparison with the mass, coupled system of five Boltzmann kinetic equations
for all stable components (with an account of co-annihilation reactions) is numerically solved.

As it is shown in detail in \cite{IJMPA_2019}, there are a set of regions in a plane of H-pion and H-sigma masses (see Fig.1), where it is possible to fix the DM relic density in
agreement with the modern data.
\begin{figure}
    \center{\includegraphics[width=6 cm]{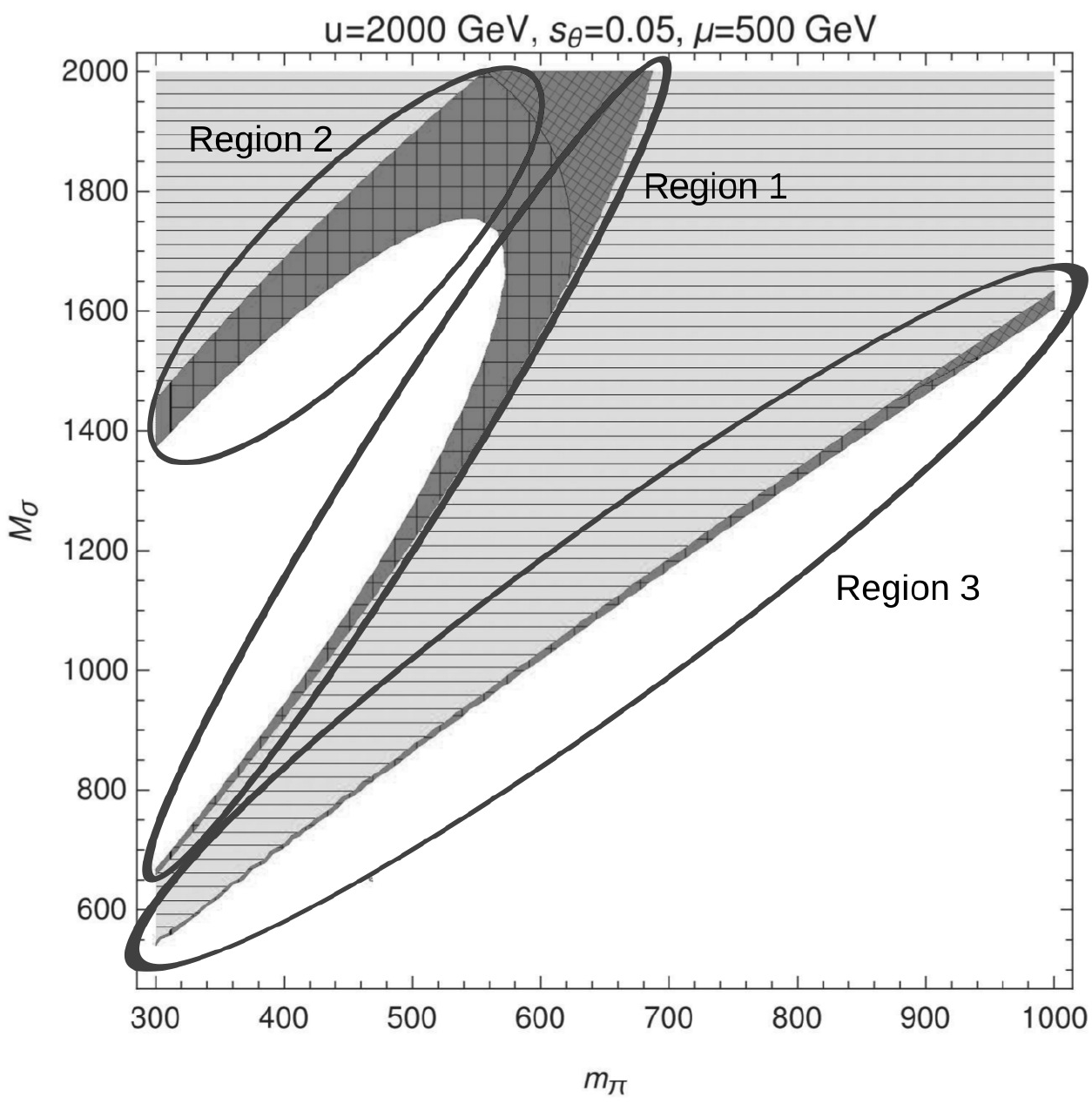}}
    \caption{Phase diagram in terms of $M_{\tilde \sigma}$ and $m_{\tilde \pi}$ which is resulted from numerical solution of the kinetic equations system.}
    \label{fig_regions}
\end{figure}
Because only H-pions interact with vector bosons, there are no regions where this component dominates in the DM density. The stable $B^0$-baryons interact with matter only via H-quark
and H-pion loops and scalar exchange channels. It is a specific feature of SU(4) vector-like model with two stable pNG states. From numerical tree level analysis there are three
allowable regions of parameters (masses):

{\bf Region 1}: here $M_{\tilde{\sigma}}>2m_{\tilde \pi^0}$ and $u \ge M_{\tilde{\sigma}} $; at small mixing, $s_{\theta}\ll 1$, and large mass of H-pions we get a reasonable value
of the relic density and a significant H-pion fraction;

{\bf Region 2}: here again $M_{\tilde{\sigma}}>2m_{\tilde \pi^0}$ and $u \ge M_{\tilde{\sigma}}$ but $m_{\tilde \pi}\approx 300-600 \,\, \mbox{GeV}$; H-pion fraction is small here,
approximately, $(10-15)\%$;

{\bf Region 3}: $M_{\tilde{\sigma}}<2m_{\tilde \pi}$ --- this region is possible for all values of parameters, but decay $\tilde \sigma \to \tilde \pi \tilde \pi$ is prohibited.
Here, H-pion fraction can be up to $40\%$ for $m_{\tilde \pi^0}\sim 1 \,\mbox{TeV}$ and small mixing between scalars.

So, possible values of H-pion mass should vary approximately in the range $(600\text{--}1200)\, \mbox{GeV}$ in agreement with recent astrophysics data. Because some hopeful results
from colliders are absent, we consider  indirect searches of DM manifestations in astrophysical data \cite{RST_Rev,Gaskins,Arcadi,Belotsky,Indirect}. Now, we come to study of
high-energy cosmic rays quasi-elastic scattering off the DM \cite{Scatt0,Scatt1,Scatt2,IJMPA_2019,Symmetry_2020}. Most simple reaction in H-color scenario is cosmic ray electrons
scattering off H-pion component via weak boson in the process $e \tilde \pi^0 \to \nu_e \tilde \pi^-$, then charged $\tilde \pi^-$ will decay as it was indicated above. In the
narrow-width approximation we get for the cross section: $\sigma(e \tilde \pi^0 \to \nu_e \tilde \pi^0 l \nu'_l ) \approx \sigma((e \tilde \pi^0 \to \nu_e \tilde \pi^-)\cdot Br( \tilde
\pi^-\to \tilde \pi^0 l \nu'_l ),$ branchings of charged hyperpion decay channels are: $Br( \tilde \pi^-\to \tilde \pi^0 e \nu'_e) \approx 0.01$ and also $Br( \tilde \pi^-\to \tilde
\pi^0 \pi^-) \approx 0.99$. Considering final charged hyperpion $\tilde \pi^-$ near its mass shell, standard light charged pion produces neutrino $e \nu_e$ and $\mu \nu_{\mu}$ with
following probabilities: $\approx 1.2\cdot 10^{-6}$ and $\approx 0.999$, correspondingly.

Then, an energetic cosmic electron produces electronic neutrino and soft secondary $e' \nu'_e$ or $\mu \nu_{\mu}$ arise from charged H-pion decays. Now, there are final states with
$Br(\tilde \pi^0 \nu_e \mu'\nu'_{\mu})\approx 0.99$ and $Br(\tilde \pi^0 \nu_e e'\nu'_e)\approx 10^{-2}$. These results are justified in the framework of the factorization approach
\cite{Kuk_Vol}.

We get that initial electron with energies in the range $E_e=(100-1000) \,\mbox{GeV}$ interacts with cross section decreases from $O(10)$ up to $O(0.1)\,\mbox{nb}$ and there is a
maximum at small angles between electron and the neutrino emitted \cite{IJMPA_2019}). In this approximation, energy of the neutrino produced is proportional to incident electron energy
and depends on the mass of the Dark Matter particle very weakly. The neutrino flux is calculated by integrating of spectrum, $dN/dE_{\nu}$, this flux depends on H-pion mass very
weakly. In the interval $(50-350)\, \mbox{GeV}$ it decreases most steeply,  and then, down to energies $\sim 1\, \mbox{TeV}$ the fall is more smoother.

We also estimate number of neutrino landings on the IceCube surface and (even with an amplification the DM density near the Galaxy center for the symmetric Einasto profile), we get
very small number of neutrino events per year: $N_{\nu} =(6-7)$. Indeed, this number can be increased for cosmic rays energy in multi-TeV region. However, the electron flux is only
small part of the cosmic rays total flux especially at energies $\geq 10^2 \, \mbox{TeV}$. As a result, we predict a very small fluxes of secondary neutrinos and, consequently, small
probability to detect such events at IceCube \cite{IJMPA_2019,Symmetry_2020}. Cosmic rays scattering off DM clusters of very high density \cite{Clumps2,Clumps3} can result in
amplifying secondary neutrino flux \cite{Clumps_amplify_1,Clumps_amplify_2}.

It seems that there is a chance to introduce the $B^0$ interaction through H-pion and/or H-quark loops, however for the scattering channels these loops are exactly zero \cite{IJMPA_2019}.
Thus, we need to consider of more complex tree diagrams, in particular, tree diagrams with the exchange of Higgs boson and its partner, $\tilde \sigma-$meson in t-channel give dominant
non-zero contribution to the process $e^-B \to \nu_e W^-B$ (see Fig.2.a). Virtual $W^- -$boson eventually decays to $l \bar \nu_l$ or into light ordinary mesons. Of course, there is
similar scattering reaction with the scalar states exchange, $e^-\tilde \pi^0 \to \nu_e W^-\tilde \pi^0$, whose amplitude is half as it is seen from the model Lagrangian. We have found,
these diagrams give dominant tree level part of cosmic particles scattering cross section, and we do not take into account small contributions from H-quark loops, $hhZ$ and other multi-scalar
vertices\cite{Symmetry_2020}.
\begin{figure}[h]
    \centering
    \begin{minipage}[h]{0.7\linewidth}
        {\includegraphics[width=0.8\linewidth]{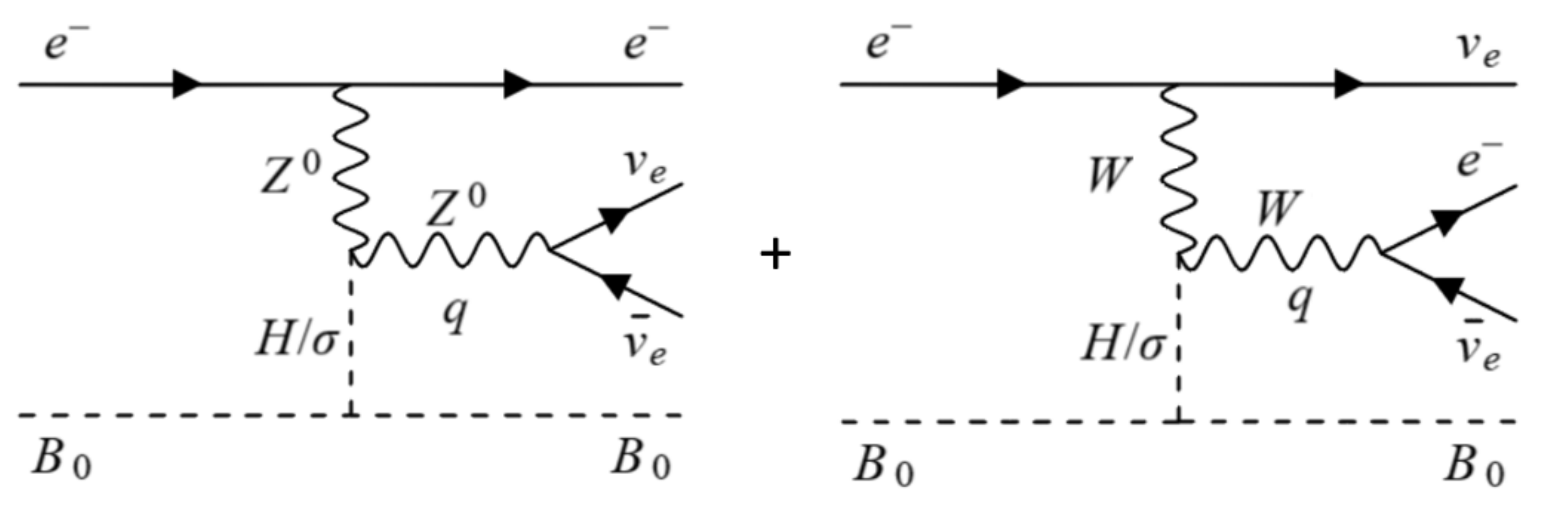} \\ a)}
    \end{minipage}
    \begin{minipage}[h]{0.7\linewidth}
        {\includegraphics[width=0.8\linewidth]{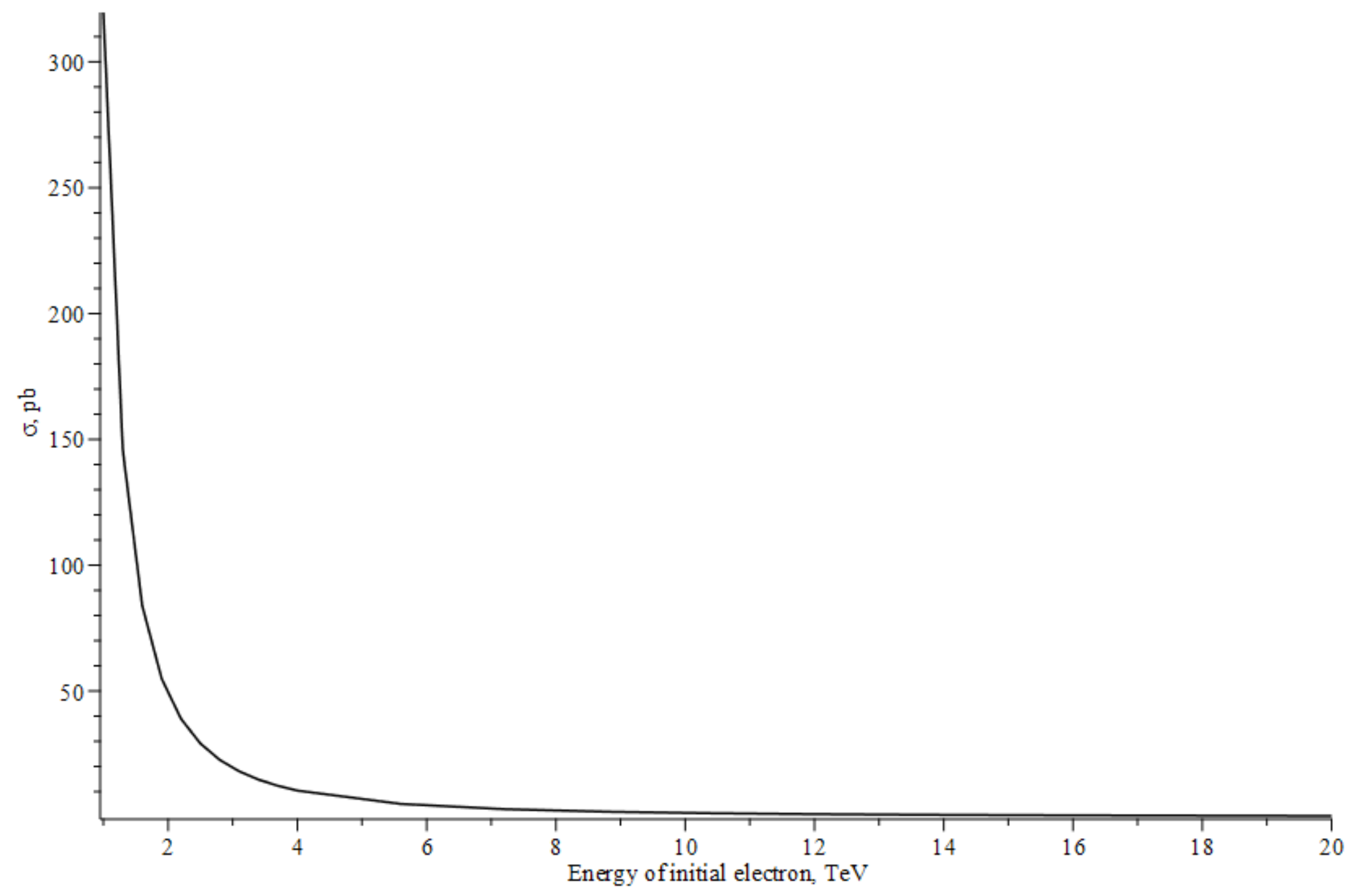} \\ b)}
    \end{minipage}
    \caption{Quasi-elastic cosmic lepton scattering with energies $(1 - 20)\, \mbox{TeV}$ off H-baryon Dark Matter component: a) necessary Feynman diagrams; b) total cross section for $m_B=1200\, \mbox{GeV}$.}
    \label{12B}
\end{figure}
To calculate total cross section of the process with final state $B^0e^-\nu \bar \nu$ or $\tilde \pi^0 e^-\nu \bar \nu$, it was used factorization method \cite{Kuk_Vol} considering
independently amplitudes squared of subprocesses with intermediate W and Z-bosons and then estimating the (negative) interference of these contributions. The approach allows to
estimate with reasonable accuracy (no worse than $\sim10\%$ due to approximate estimation of the interference) cross section of an "averaged" process where final electron and neutrinos
are produced by different vertices, $W \to l \nu_l$ and $Z\to \nu_l \bar \nu_l$.

So, we get a reasonable evaluation of total cross section (Fig.2b) and can estimate also the possibility to detect at IceCube the neutrino signal producing by the process of electron
scattering off the DM. In this calculation, we restrict ourselves those phase space regions which do not include an acceleration of initial DM particle.  In other words, final DM
components are slow and nearly all energy of the incident lepton is distributed between three final massless leptons (electron and pair of neutrinos). Of course, for neutrino scattered
via W-vertex cross section is the same. The secondary neutrino fluxes calculated are very small in comparison with expected neutrino fluxes from AGN which can be $\sim 10^5$
$cm^{-2}s^{-1}sr^{-1}$. Atmospheric neutrino fluxes with neutrino energies $\leq 2 $TeV are also much larger \cite{NeutrinoFlux1,NeutrinoFlux2,NeutrinoFluxAtm}: $\sim 10^{-10}-10^{-9}$
$cm^{-2}sr^{-1}s^{-1}$. Namely, we get the secondary
 neutrino flux resulted from the cosmic electrons scattering is $\sim 10^{-19}-10^{-22}$ $cm^{-2}sr^{-1}s^{-1}$ \cite{Symmetry_2020}.

Interestingly to note a specific scattering process when high-energy intergalactic neutrino interact with the DM via neutral Z-vertex as
$\nu_l+\mbox{DM} \to \bar \nu_l+Z^* +\mbox{DM} \to \bar \nu_l+ \nu_k\bar \nu_k+\mbox{DM}$. Then three secondary neutrinos are produced and can be accompanied with the accelerated DM particle.
 This process can be informative especially because both of these neutral objects can be messengers from regions of high DM density --- regions near AGN or from possible DM inhomogeneities
 of some other nature ---and early epoch of the Universe \cite{Neutrino_Early}. Work on analysis of such reactions is in progress.

Thus, there are some points which are important for study of the cosmic rays scattering off the DM. Independently of the SM extension, processes with scalar exchanges results in a strong
dependence of the cross section on the DM particle mass giving dominant part of the total ctoss section. In H-color scenario, increasing of the DM of $10\,\%$ provides the cross section
grow up to $50\,\%$. The opening of channels with scalar exchanges allows to consider an additional ways to produce secondary high-energy leptons and photons by ultra high-energy cosmic
rays (UHECR) scattering off the DM.

\section{High energy protons and the Dark Matter particles acceleration}

Remind that expected number of neutrino events is too small to be measured in experiments at modern neutrino observatories.
The weakness of the signal is also resulted from effective bremsstrahlung of electrons and the smallness of electron fraction in cosmic rays, $\sim 1 \, \%$. Therefore, they are not so
good probe for the DM structure; only if there are sharply non-homogeneous spatial distribution of hidden mass, the signal of production of energetic neutrino by cosmic electrons can be
detected. It is an important reason to study inelastic scattering of cosmic protons, because they are more energetic and have a much larger flux.

The possibility to accelerate light DM particles due to scattering of high-energy cosmic rays off the DM was recently supposed and numerically analyzed in a numerous papers
\cite{Acc1,Acc2,Acc3,Acc4,Acc5,Acc10CR,AccDM}. In \cite{Acc_Boosted} kinetic energy of DM particle which was initially at rest and then has been accelerated by high-energy cosmic ray
particle, was calculated assuming the scattering reaction is elastic and isotropic. This simplified analysis of the DM acceleration was used for the light DM candidates. It seems,
however, in the approximation of elastic reaction in the CMS we can use the same simple formula from \cite{Acc_Boosted} to estimate possibility of boosting for heavy DM objects.
Namely, we have
\begin{align}
T_{DM} = \frac{1}{2}(T_{CR}^2+ 2mT_{CR})(T_{CR}+M/2)^{-1}\cdot(1+\cos\theta),
\end{align}
 where m and M - masses of cosmic ray and DM particles, correspondingly, $T_{DM}$ and $T_{CR}$ - kinetic energies of DM particle after the scattering and projectile (cosmic ray particle),
  $\theta$ - angle of scattering in CMS (here we consider $M>>m$). Further, for $T_{CR}>>M$ we get an estimation $T_{DM} \sim T_{CR}$. So, heavy DM objects with masses $\sim 1\, \mbox{TeV}$,
  as it takes place in H-color scenario, can be effectively accelerated by cosmic ray protons of high energies $\sim 10^2 \,\mbox{TeV}$. Cosmic rays with such energies can be generated near
  AGN, in particular.
 In other words, fast protons from blazar’s jets can interact with heavy DM particles from halo having the largest density near AGN. In the framework of the H-color scenario, a significant
 part of protons energy can be transferred due to charged current to heavy H-pion and to both DM component in the diagrams involving scalar exchange as it takes place for leptons scattering
 off H-baryon component (see Fig.2a).

As it is seen from Fig.3a,b, if an initial proton with energy in the range $50 - 200\, \mbox{TeV}$ interacts with neutral H- pion which is nearly at rest, cross section of the
scattering process where final charged H-pion is produced with energies $(40-50)\, \mbox{TeV}$ is $\sim (10-15)\, \mbox{pb}$. Certainly, secondary charged H-pion predominantly decays
as $\tilde \pi^{\pm} \to \tilde \pi^o \pi^{\pm}$ with the width $\Gamma\approx 3\cdot 10^{-15} \, \mbox{GeV}$. So, besides neutral H-pion there appear secondary muon and muonic
anti-neutrino in the final state.

Thus, in the deep inelastic reaction the main charged component of UHECR i.e. protons, can transform in part into flux of high energy neutrino and leptons accompanied with accelerated DM
particle. From our estimations it follows that $\sim (10-25)\%$ of the proton energy is transferred to heavy neutral component of the DM with cross section $\approx (10-100) \, \mbox{pb}$.
When the UHECR scatter off B-component, total cross section is of the same order but there can appear additional neutrinos generated by decay of intermediate vector bosons.

It should be noted, to get an estimation of these cross sections we have used a very simple model of quark distribution function in the proton:
$q(x) \approx A\cdot x^{\alpha}\cdot(1-x)^{\beta}$. In other words, we used very simplified an analytical expression for pdf's (see also Refs.~\cite{PDF_1,PDF_2})
because at these high energies we do not know an exact form of pdf's and only try to get some reasonable evaluation of the cross section. We assume, an error in these estimations cannot
be more than in one order.
\begin{figure}[h]
    \centering
    \begin{minipage}[h]{0.7\linewidth}
                {\includegraphics[width=0.9\linewidth]{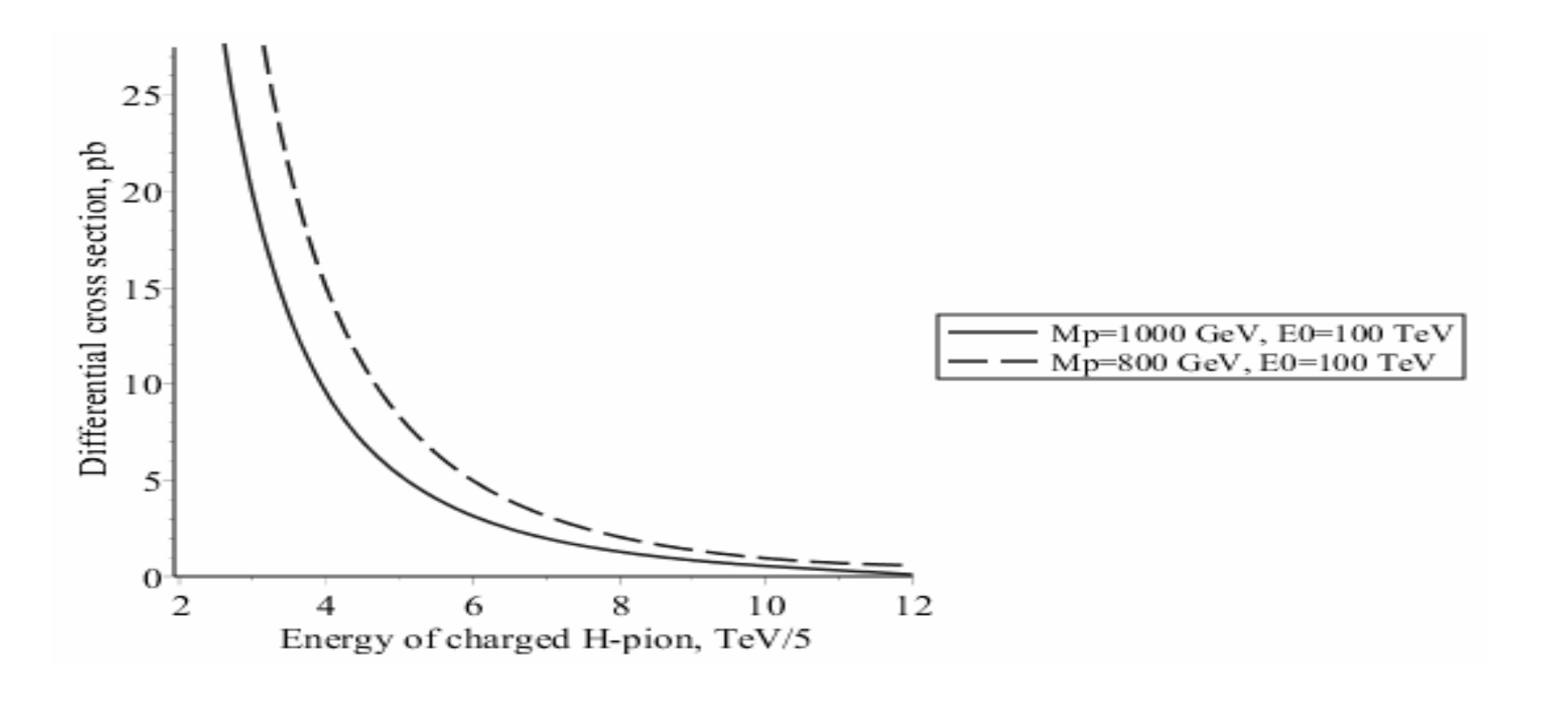} \\ a)}
    \end{minipage}
    \begin{minipage}[h]{0.7\linewidth}
                {\includegraphics[width=0.9\linewidth]{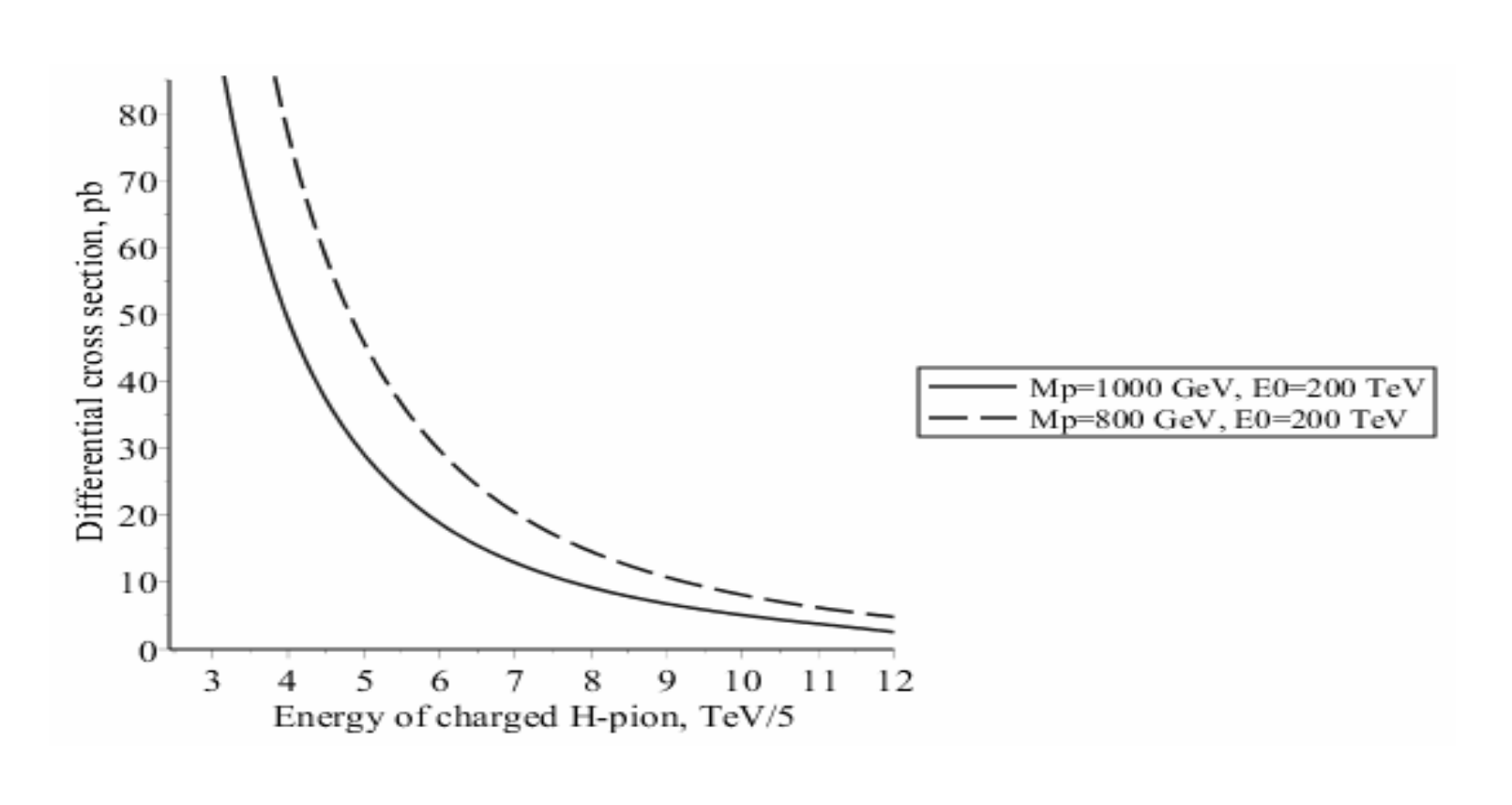} \\ b)}
    \end{minipage}
    \caption{Differential cross section for cosmic ray proton scattering off H-pion DM component: a) initial proton energy $100 \, \mbox{TeV}$; b) initial proton energy $200 \, \mbox{TeV}$.}
    \label{12B}
\end{figure}
Indeed, cross section of the scattering with the DM particle acceleration is small, but in these rare events heavy DM particles can accelerate significantly and pass away from the DM
halo. Then, they move like neutrino but slower, and keep nearly constant direction due to weak interactions with the matter. So, this rare process when the charge component of cosmic
rays can be ruined in the deep inelastic reaction and as a result neutral DM particle moves like a neutrino towards the Earth. Remind that above considered high-energy electron
scattering off the DM can also accelerate the DM but in quasi-elastic process high energy neutrino are generated with more probability.

Thus, from this brief description of some processes of scattering of high-energy cosmic ray particles off the DM we can conclude that these reactions can enrich the cosmic rays
composition with boosted heavy neutral DM particles \cite{CRComp}. At energies of these projectiles $\sim (10-100)\, \mbox{TeV}$ cross sections of their interactions with nucleons and
nuclei, $\sim(10^{-34} - 10^{-37})$ $cm^2$, are compared with cross sections of neutrino-nucleons scattering. In this deep-inelastic process nucleons or nuclei are transformed a
multiparticle final states consisting of charged leptons, photons and neutrino. Additional neutrinos are generated by the charged H-pion decay and in the processes with resonant decay
of Z-boson. So, the accelerated neutral DM components can produce rare events - specific types of extended air showers (EAS), which can be separated in the atmosphere from other types
of showers \cite{EUSO,EAS_Rev}.

It is known that as usually cosmic rays generate a shower of secondary particles which are mainly muons, electrons and photons. They go to ground detectors and can be fixed as measured
signals registering also due to fluorescence and Cherenkov light, and radio emission generated by charged component, electrons, in atmosphere of the Earth. It seems, such type of shower
is similar to neutrino induced shower and its initial point also should be deeply in atmosphere, however, the neutral DM particle can not disappear from the EAS composition and will
interact with the ground detector producing some radiation from secondary electrons or from excited nuclei in the detector. The DM showers, as they generated by intergalactic DM objects
which were accelerated by UHECR or AGN jets from halo of other galaxies, or DM particles boosted  from halo of our Galaxy by intergalactic UHECR do not have to be mostly inclined or nearly
horizontal. It is supposed, these accelerated DM components and EAS produced by them should be distributed more or less isotropic. May be, the EAS axis can be connected with direction to
some blazar, as it was found for some very high-energy neutrino events at IceCube.

So, we can conclude that EAS from heavy DM particles are distinguished from EAS generated by protons or neutrino because in the former event the shower contains in his composition
neutral stable object up to the final moment when this fast DM particle scattered on nucleon in the detector (see also \cite{DM_Search} and references therein). In contrary, in
composition of EAS which was induced by neutrino or protons (or light nuclei), there is no any heavy stable particles, only leptons, photons and neutrino are detected as final states.
Note also that interaction of DM component with nucleons in detector should have specific signature: the scattering in charged current channel is accompanied with creation and
following decay of charged H-pion, so, the event can be seen due to charged lepton bremsstrahlung. We hope, observing and measurement of characteristics new types of EAS containing
heavy  neutral stable particle will be possible at modern complex LHAASO \cite{LHAASO}, in other words, the DM candidates can manifest itself in a specific types of EAS.

\section{Conclusions and Discussion}
It is known, hadrons, leptons, photons with energies $E \geq 10^7$ TeV  cannot reach the Earth because they interact with $\gamma$-bkg and loss the energy. High-energy photons
with $E \geq 10$ TeV also practically cannot reach the Earth due to interactions with $\gamma$-bkg of various wave lengths --- electron-positron pairs creation decreases photon
energies below 10 TeV.
But intergalactic high energy neutrino can move to the Earth being generated, for example, in blazars jets. Neutrinos being produced in decays of high-energy hadrons and in reactions
of scattering and conserve their energy on the way from remote sources – at cosmological distances in 10-100 Mps or more.
Certainly, neutrinos can be also produced by supermassive X-particles decay and in virtual Z-boson resonant transition to neutrino or hadronic pairs or from resonant generation of
lepton + neutrino pair or hadronic pairs from virtual W-boson. And when neutrinos reach the Earth's atmosphere, they can produce Extended Atmospheric Showers (EAS) with high portion
of neutrino energy despite of small interaction cross section (which, however, increases with energy).

However, as we see, there is a possibility to accelerate (heavy) neutral DM particles which also can move from the distant sources, as the neutrino.
EAS generated by neutrino can be successfully discriminated from other types of events (from EAS induced by fast protons, for instance) because they are produced at large depth in
atmosphere and are mostly strongly inclined or they are even nearly horizontal. It is an important "fingerprint" for the EAS detection at modern complex , LHAASO. We assume, heavy
accelerated DM particle also would produce EAS more deeply than ordinary cosmic rays, mimicking, in fact, neutrino event but with different secondary particles spectrum and total
energy release in the process. So, if the DM particles entrance into atmosphere with sufficiently high energy, they can produce specific EAS similar to neutrino-induced ones only
in some aspects.

There are known a number of neutrino events with energies up to $E \approx 10^7$ TeV which were registered at IceCube. The source of such super high-energy neutrinos is still unknown,
and maybe resonance  at $q^2 = M_Z^2$ can contribute to high-energy neutrino creation when high-energy proton interact with the DM particles. In this process some part of proton energy
transferred to multiple secondary decaying mesons and neutral accelerated DM.  High-energy intergalactic neutrino scattered by DM in halo can transfer its energy to secondary leptons and
accelerated neutral DM object. This event can be detected as correlated EAS produced by neutral objects with energies up to $\sim 10^3 \, \mbox{TeV}$. In this range of energies the
atmospheric bkg should be small.
In fact, cross section of neutrino-DM interaction is $\sim 10^2 \, \mbox{pb}$, it is much lower than
cross section for annihilation of high-energy neutrino with relic neutrino which is $\approx 10$ nb
but such type events, in principle, can be registered at IceCube and LHAASO as correlated EAS.

Note, cosmic ray proton scattering off the DM can give rise to increasing of positrons number --- they are products of eventually decay of (positively) charged secondary hyperpions.
Energy spectrum of these secondary  positrons are determined by energies of cosmic rays primaries , masses of the DM candidates, type of the scattering reaction and kinematics of charged
H-pion decay. This process does not considered in detail yet.

If it wold be found some increasing of secondary particles (neutrino and/or leptons) flux and the number of  detected events from some fixed direction, it can follow from the UHECR
scattering off the DM clumps. In other words, to study the DM space distribution, an analysis of EAS induced by (different) neutral objects, their correlation together with measurements
of secondary neutrino spectrum and the number of events, can be used.

Moreover, there are some other features of the hypercolor two-component DM scenario, namely, at high energies inelastic reactions with the exciting of higher states of the pNG unstable
H-hadrons can occur. Arising and decays of these excited states can be manifested as heavy H-hadron jets that eventually decay to neutral stable DM particles accompanied with photons,
leptons and decaying standard light mesons. To study these processes we should know (or suppose) the mass spectrum of unstable H-hadrons, their possible decay channels and widths.
In any case, the DM two-component structure can be seen studying of correlations in the set of quantitative and qualitative results in vector and scalar UHECR scattering channels.

\section*{Acknowledgements}

The work was supported by grant of the Russian Science Foundation (Project No-18-12-00213)

\end{document}